\documentclass[10pt, pra,aps,twocolumn,showkeys]{revtex4-2}
\usepackage[latin1,utf8]{inputenc} 
\usepackage{amssymb}
\usepackage{amsfonts} 
\usepackage{graphicx} 
\usepackage{caption}
\usepackage{enumerate}
\usepackage{svg}
\usepackage{float}
\usepackage{amsthm} 
\usepackage{amsmath} 
\usepackage{bm}
\usepackage{epsfig}
\usepackage{hyperref}
\usepackage{color}
\usepackage{bbold}
 \usepackage{bbm}
\usepackage[T1]{fontenc}
\usepackage{young}
\usepackage{youngtab}
\usepackage{xcolor}
\usepackage{braket}
\usepackage{lipsum}
\usepackage{comment}
\usepackage{orcidlink}

\def \bc {\begin{center}}
\def \ec {\end{center}}
\def \bi {\begin{itemize}}
\def \ei {\end{itemize}}

\def \ba {\begin{array}}
\def \ea {\end{array}}

\def \bea {\begin{eqnarray}}
\def \eea {\end{eqnarray}}

\def \be {\begin{equation}}
\def \ee {\end{equation}}

\newcommand{\la}{\langle}
\newcommand{\ra}{\rangle}

\def \um {\frac{1}{2}}

\def\tr {\mathrm{tr}}

\def\bc {\bar{\beta}}

\def\x {^\dagger}
\def\Dc{\mathcal{D}}

\def\2cat{{\scriptstyle\mathrm{2CAT}}}
\def\3cat{{\scriptstyle\mathrm{3CAT}}}

\begin{document}

\title{Electron Shuttle Waiting Times for Electric Field Sensing
}

\author{Alberto Mayorgas\,\orcidlink{0000-0002-7071-2025}}
\email{alberto.mayorgasreyes@ceu.es}
\affiliation{Universidad CEU Fernando III, CEU Universities,
Escuela Polit\'ecnica Superior, Glorieta Cardenal Herrera Oria,
41930 Seville, Spain}

\author{Alberto López-García\,\orcidlink{0009-0001-0023-9850}}
\affiliation{Universidad Polit\'ecnica de Cartagena member of European Uniersity of Technology EUT+,
Research Group of Quantum Technologies, \'Area de F\'isica Aplicada,
Cartagena E-30202, Spain}

\author{Enamul Haque\,\orcidlink{0000-0002-5291-7203}}
\affiliation{Universidad Polit\'ecnica de Cartagena member of European Uniersity of Technology EUT+,
Research Group of Quantum Technologies, \'Area de F\'isica Aplicada,
Cartagena E-30202, Spain}

\author{Javier Cerrillo\,\orcidlink{0000-0001-8372-9953}}
\email{javier.cerrillo@upct.es}
\affiliation{Universidad Polit\'ecnica de Cartagena member of European Uniersity of Technology EUT+,
Research Group of Quantum Technologies, \'Area de F\'isica Aplicada,
Cartagena E-30202, Spain}

\date{\today}

\begin{abstract}
	\vspace{0.4cm}
	\section*{Abstract}
We explore the use of waiting-time statistics in a quantum electron shuttle for electric-field sensing. Electron shuttles convert nanomechanical motion into charge transport, showing a noise-broadened crossover between stochastic tunneling and mechanically assisted charge transfer. This allows investigation of how transport fluctuations encode electromechanical parameters. Using a single-level quantum shuttle in strong-Coulomb-blockade and high-bias regimes with a Markovian quantum master equation, we analyze stationary dynamics in phase space and waiting time distributions. By estimating the electromechanical coupling, proportional to the electric field, we evaluate the classical Fisher information in waiting times and compare it with the quantum Fisher information of the stationary state. We relate the metrological response to mean waiting time, variance, and Fano factor. Our results show that the crossover from tunneling to shuttling is characterized by enhanced fluctuations and increased parameter sensitivity, leading to a pronounced enhancement of the Fisher information.
\end{abstract}

%
%
%
%
%


\maketitle


\section{Introduction}
Nanoelectromechanical systems offer a natural and unique platform for investigating the interplay between charge-carrier dynamics and mechanical motion. In such a system, charge-carrier tunneling induces a force on a movable mechanical element, while the element's displacement, in turn, modulates the tunneling amplitudes. A noteworthy example of such a system is the electron shuttle, in which a small island or quantum dot oscillates between two leads and transfers charge carriers between the source and the leads, or vice versa, through repeated loading and unloading events \cite{armour2002transport,fedorets2003quantum, fedorets2004quantum, armour2004classical}. The shuttle mechanism was originally proposed as a self-sustained charge-transfer process in Coulomb-blockade nanostructured systems, where mechanical motion cooperates with sequential tunneling \cite{bachtold2022mesoscopic}. Consequently, this coupled motion generates a current whose magnitude is determined by the oscillator’s frequency \cite{gorelik1998shuttle}. Such an electron transport cycle cannot be simplified to a standard single-electron transistor nor an isolated mechanical oscillator. While electromechanical force can induce a relatively sharp transition from the tunneling regime to the self-oscillatory shuttling regime in the classical mean-field picture, oscillator efficiency in the quantum regime depends on quantum fluctuations and environmental noise. In the quantum regime, these fluctuations turn the sharp classical instability into a smoother crossover \cite{PhysRevLett.80.4526,Novotny2003, fedorets2003quantum, fedorets2004quantum, armour2004classical}. In this regime, we can describe the stationary state in phase space using quasi-probability distributions, e.g., the Wigner or Husimi function \cite{novotny2003quantum}. Thus, the quantum shuttle provides an intriguing and useful example of an open quantum system in which transport, dissipation, and mechanical motion are strongly coupled. Depending on the coupling, the stationary dynamics of an electron shuttle fall into three regimes. In the tunneling regime, charge transfer occurs through stochastic tunneling, and the mechanical part remains localized near equilibrium. In the shuttling regime, charge and mechanical motion are coupled, with electrons transferring in sync with the oscillator. Between these regimes, a crossover occurs, with coexisting incoherent tunneling and mechanically assisted transport. Although a sharp classical transition exists, quantum and thermal fluctuations turn it into a crossover region \cite{novotny2003quantum,PhysRevB.74.014303}.

Transport noise is key to identifying these regimes. Phase-space distributions, current cumulants, and electron waiting times reveal this dynamical shift \cite{PhysRevB.69.245409, blanter2000shot, levitov1996electron, emary2007frequency, flindt2008counting}. In particular, zero-frequency current noise and the Fano factor quantify charge-transfer fluctuations \cite{zhao2025non,Novotny2004, PhysRevB.69.245409, blanter2000shot, levitov1996electron, emary2007frequency, flindt2008counting}, while waiting time distributions (WTD) capture the timing between tunneling events \cite{Brandes2008}. These measures provide information beyond the stationary current, distinguishing nearly Poissonian tunneling from regular timing tied to the mechanical cycle \cite{albert2012electron, thomas2013electron, haack2014distributions, stegmann2021electron}. Waiting-time statistics help characterize single-electron sources, molecular junctions, and mesoscopic conductors with internal dynamics \cite{zhao2025non, Cerrillo2022,Brandes2008}. Interestingly, the same transport record may also be interpreted as a metrological signal \cite{paris2009quantum, degen2017quantum, alipour2014quantum, gammelmark2014fisher}. Device parameters that are not directly measurable can be inferred from their influence on the observed tunneling statistics. However, the information obtained through this estimation protocol depends on the measurement performed. For instance, the classical Fisher information quantifies the sensitivity of a specified probability distribution—in this case the electron waiting time distribution—to variations in the unknown parameter \cite{PhysRevA.94.042322, ma2019improving}. Hence, it evaluates the performance of a particular electron-counting strategy. In contrast, quantum Fisher information is independent of any specific measurement and determines the maximum classical Fisher information attainable by optimizing all measurements on the relevant quantum state \cite{PhysRevLett.72.3439,Wolf2019}. The distinction is crucial: a convenient transport observable may be experimentally accessible while still disregarding information embedded in other electronic or mechanical degrees of freedom \cite{PhysRevA.94.042322, ma2019improving, garrahan2010thermodynamics, radaelli2026parameter}. This raises two questions: how much parameter information encoded in the shuttle can be extracted from tunneling-event timing, and how does sensitivity relate to transport noise during the tunneling-shuttle crossover? Although noise, waiting-time statistics, and shuttle dynamics have been studied extensively, their relationship to measurement-dependent Fisher information has received less attention.

Here, we address these questions about a quantum shuttle described by a Markovian master equation in the regimes of strong Coulomb blockade and high bias. Using an electron shuttle, we determine the stationary island-oscillator state and its Husimi representation to identify tunneling, crossover, and shuttling regimes. The bias-generated electric field exerts an occupation-dependent force on the oscillator. We then calculate the distribution of waiting times between successive tunneling events and extract its mean, variance, and Fano factor. These quantities reveal how the temporal structure of electron transfer changes as electromechanical coupling increases. We subsequently treat the electromechanical coupling as the parameter to be estimated. The classical Fisher information is obtained from the complete waiting time distribution, rather than only from its first moments, and is compared with the quantum Fisher information of the stationary state. Within the resource convention adopted in this work, the waiting-time Fisher information remains below the quantum Fisher information over the considered parameter range. The difference quantifies the information encoded in the shuttle state but not accessed by the selected tunneling-time measurement. We further find that both enhanced waiting-time noise and enhanced sensitivity occur in the crossover region, although their maxima do not generally coincide. Overall, our results connect three descriptions of the electron shuttle that are often treated separately: phase-space dynamics, time-resolved transport noise, and quantum parameter estimation.

\section{Model}\label{sec2}
The model consists of a quantum dot or island with a single electronic level moving between two leads/electrodes (source and drain). This is equivalent to a quantum dot single electron transistor (SET) in which the island oscillates and modulates the tunnel conductance, without including a charging gate for the island. A bias voltage between the leads causes an electron to tunnel from the source onto the island, which moves towards the drain sopporting the electron to tunnel again and creating a current. We use the strong Coulomb blockade regime so that there is only one electronic level free in the island. The Hamiltonian of the entire system is schematically presented as
\be\label{HamTotal}
H_{\text{T}}=H_{\text{S}}+H_{\text{L}}+H_{\text{tun}}+H_{\text{B}}\,,
\ee

The first component is the island-oscillator Hamiltonian $H_{\text{S}}$. The electronic single state on the dot is described by the fermionic operators $c,c\x$, and the vibrational degree of freedom by the bosonic operators $a,a\x$, with their respective anticommutation and commutation relations. The first Hamiltonian reads
\be\label{Hamiltonian}
H_{\text{S}}=\varepsilon_\text{I} c\x c + \hbar\nu a\x a - \hbar\chi c\x c (a\x + a)\,,
\ee
describing a two-level system (single electronic level) of energy $\varepsilon_\text{I}$, a quantum harmonic oscillator of frequency $\nu$ and a coupling term with strength $\chi$ proportional to the electric field
due to the bias voltage between the left and right leads, that is, the electric field seen by an electron on the dot.

\begin{figure}[h]
	\centering
	\includegraphics[width=0.48\textwidth]{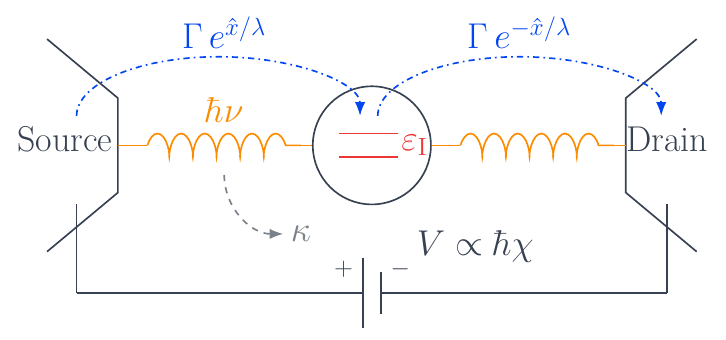}
    \caption{Schematic representation of the electron shuttle. The central island (quantum dot) is modeled as a two-level system with level spacing $\varepsilon_{\mathrm{I}}$ (red) and moves in a harmonic potential with energy quantum $\hbar\nu$, represented by the orange springs. The blue dash-dotted arrows indicate position-dependent electron tunnelling from the source to the island and from the island to the drain, with rates $\Gamma e^{\hat{x}/\lambda}$ and $\Gamma e^{-\hat{x}/\lambda}$, respectively, where $\hat{x}$ is the island displacement and $\lambda$ is the tunnelling length. The applied bias $V$ controls the island--oscillator coupling strength $\chi$, while the gray dashed arrow denotes mechanical damping at rate $\kappa$ due to the coupling of the oscillator to a thermal bath.}
	\label{Fig:shuttle}
\end{figure}

We glance over the rest of the terms in the Hamiltonian, for a detailed discussion see the references \cite{Milburn2006,Zhao2025} and Appendix \ref{App:Approximations}. The second component $H_{\text{L}}$ characterizes the leads, the electrostatic
energy of an ensemble of non-interacting electrons, which are in local equilibrium and described by the Fermi distribution. The third term $H_{\text{tun}}$ is the tunnel coupling between the island and the leads. The tunneling amplitudes are proportional to the exponential $e^{\pm \hat{x}/\lambda}$, which is a function of the tunneling length $\lambda$ and the distance of the island to the leads $\hat{x}=x_0(a+a\x)$, where $x_0=\sqrt{\hbar/2 m \nu}$ is the zero point motion of the oscillator and $m$ its mass. The fourth term $H_\text{B}$ includes the coupling of the oscillator to a dissipative Ohmic heat bath in the rotating wave approximation (weak bilinear interaction) \cite{Zoller,Weiss,Petruccione}. This produces damping and thermal noise in the mechanical system, and set a boundary in the motion.

The master equation (ME) is derived tracing the degrees of freedom of the leads and the heat bath, hence the equation contains only the island-oscillator subsystem \cite{Milburn2006,Novotny2003},
\bea\label{ME}
\frac{d}{dt}\rho&=&-i\nu\left[a\x a,\rho \right]+i\chi\left[c\x c(a\x+a),\rho \right]+ \kappa \,\Dc\left[a\right]\rho \\
&+& \Gamma\, \Dc\left[c\,e^{ x_0(a\x + a)/\lambda }\right]\rho+\Gamma \,\Dc\left[c\x e^{-  x_0(a\x + a)/\lambda }\right]\rho, \nonumber
\eea
where $\Gamma$ is the rate of electron tunneling (equal for source-island and island-drain), $\kappa$ the damping rate, and the dissipator is defined as \cite{Petruccione}
\be
\Dc\left[A\right]\rho=A\rho A\x - \um \{A\x A, \rho \},
\ee
with the $\{A,B\}=AB+BA$ the anticommutator. The density matrix has a size $2N\times 2N$ due to the tensor product Hilbert space of the island-oscillator, where $N$ is the truncated size of the oscillator basis. All the frequencies and rates of the ME $\omega_{\text{I}},\nu,\chi,\Gamma,\lambda,\kappa$ are considered real and constant parameters, according to the approximations considered below.

The master equation \eqref{ME} is calculated in the Born-Markov approximation, that is the mechanical damping caused by the heat bath is addressed within the standard weak coupling theory and with a short bath correlation time \cite{Petruccione,DonariniThesis,Milburn2006}. Both estimations are combined with the rotating wave approximation to produce a quantum optical ME for the damping term. See \cite{Zhao2025} for a non-Markovian version of the ME. The bath is also in thermal equilibrium, described by the Bose-Einstein distribution as a function of the oscillator frequency and bath temperature, which is approximated to 0 \cite{Novotny2004}. This is why there is no $\Dc\left[a\x\right]\rho$ term in the ME \eqref{ME}. See Appendix \ref{App:Approximations} for more details about the approximations considered in this model that lead to the ME \eqref{ME}.

It is known that three distinct regimes are distinguished for the electron shuttle \cite{Cerrillo2022,Milburn2006,Novotny2003}:

\begin{enumerate}[I)]
    \item \textit{Tunneling (SET)}: There are two ways of damping the shuttle: 1) large mechanical damping $\kappa$ (underdamping) or 2) large jump $\Gamma$ compared to $\nu$ (overdamping). The WTD profile at 1) is a long period $t\gg 2\pi/\nu$ envelope (Gamma distribution) with small oscillations on top due to quantum noise effects, while at 2) is peaked at a time  $t\ll 2\pi/\nu$ (tunneling, characteristic for a single electron transistor). We focus on damping 1) for convenience.
    \item \textit{Shuttle}: The tunneling rate is similar to the oscillator frequency $\Gamma\simeq \nu$, so the WTD profile is a single oscillation that peaked at $t=2\pi/\nu$.
    \item \textit{Crossover}: Intermediate regime between tunneling and shuttle, where the WTD envelope shape start changing from Gamma-like to a exponential-like, and finally to a single peak (shuttle).
\end{enumerate}

In our case, we propose a scheme where the coupling strength $\chi$ is chosen as free parameter to move along the three regimes. In practice, this approach is equivalent to tune the system with a variable voltage $V\propto \hbar \chi$, which is straightforward in an experimental set up compared to, e.g., manipulate the tunneling rate $\Gamma$ or the damping $\kappa$.


\section{Results}


The stationary solution of the master equation \eqref{ME} is calculated using the vectorization procedure, which transforms the Lindblad form of the master equation $\mathcal{L}\rho_0=0$ into a system of linear equations $4N^2\times 4N^2$. Go to  Appendix \ref{App:Vectorization} for more details about the vectorization process and the challenges of finding the solution $\rho_0$ of the system. For simplicity along this work, we will use a set of parameters $\Gamma=0.01\nu$, $\kappa=0.05\nu$, $\lambda=2x_0$ and an oscillator basis of size $N=50$. The coupling $\chi\in[0.1,0.6]$ varies in a 100 element interval. For these parameters, the convergence of the stationary solution calculation is ensured since the residual remains low $\text{max}\|\mathcal{L}\rho_0\|<3.09\times10^{-14}$.

\begin{figure}[h]
	\centering
	\includegraphics[width=0.5\textwidth]{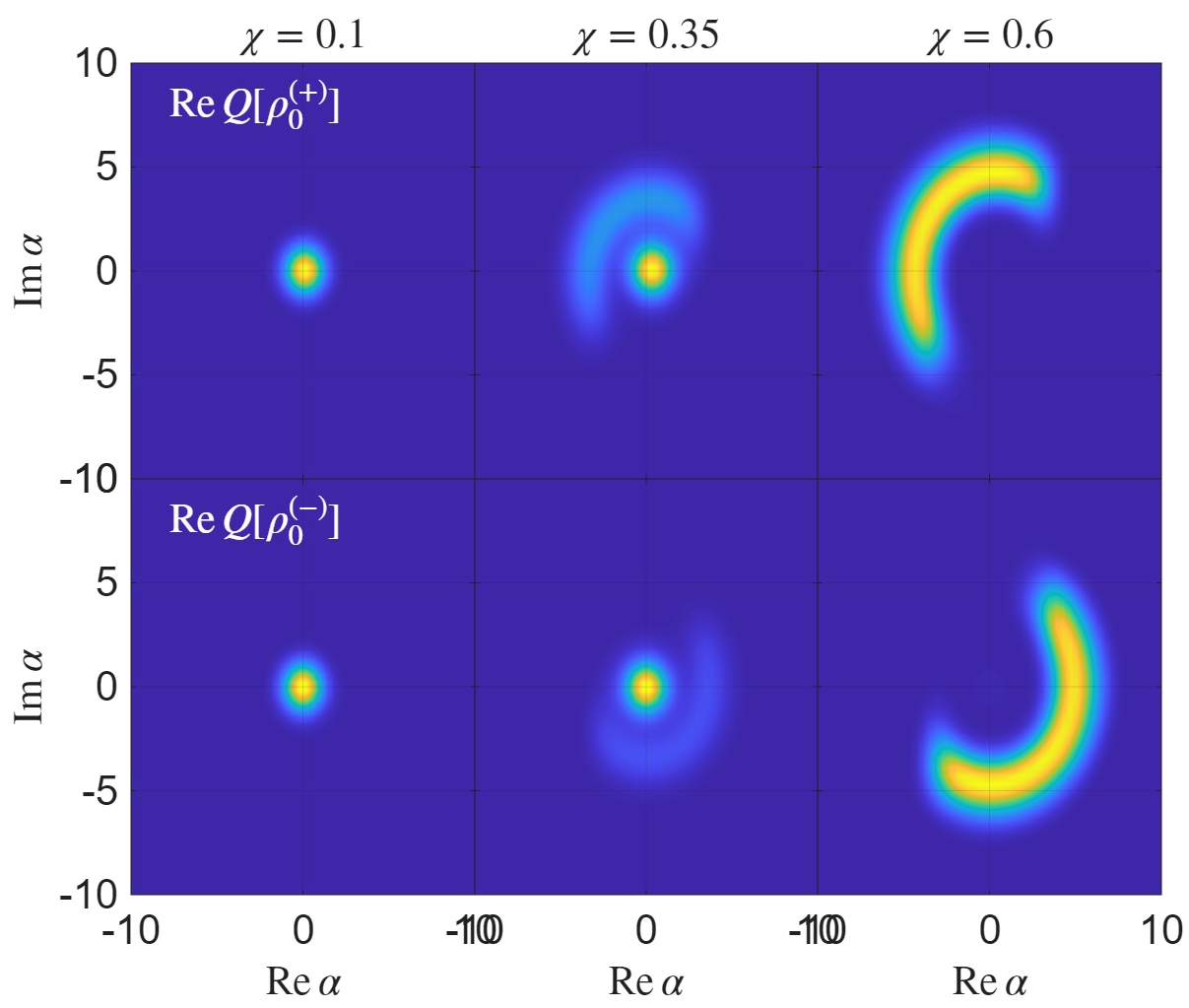}
	\caption{Husimi function of the stationary solution in the electronic diagonal elements subspaces $\rho_0^{(+)}=\bra{1}\rho_0\ket{1}$ (top row) and $\rho_0^{(-)}=\bra{0}\rho_0\ket{0}$ (bottom row), for three values of the electric field $\chi=0.1\nu,0.35\nu,0.6\nu$ in the tunneling, crossover and shuttle regimes  respectively (left to right columns). The axis represent the real and imaginary part of the phase space parameter $\alpha$. The parameters used are $\Gamma=0.01\nu$, $\kappa=0.05\nu$, $\lambda=2x_0$ and an oscillator basis of size $N=50$. }
	\label{Fig:Husimi}
\end{figure}

An useful technique to characterize the stationary solution, and distinguish different regimes, is the Wigner quasiprobability function \cite{Novotny2003}. Nevertheless, a simpler phase space representation is given by the Husimi function, whose computation is more straightforward than the Wigner and reproduces the same effect. The Husimi function  $Q(\alpha)=\bra\alpha\rho\ket{\alpha}$ is defined in terms of the harmonic oscillator (Heisenberg-Weyl) coherent states $\ket{\alpha}=e^{-|\alpha|^2/2}\sum_{n=0}^N \tfrac{\alpha^n}{\sqrt{n!}}\ket{n}$ in the truncated Fock basis. We represent the real part of the Husimi function in Figure \ref{Fig:Husimi}, separating between electronic diagonal elements $\rho_0^{(+)}=\bra{1}\rho_0\ket{1}$ (top row) and $\rho_0^{(-)}=\bra{0}\rho_0\ket{0}$ (bottom row), where $\ket{1}\bra{1}=c\x c$ and $\ket{0}\bra{0}=cc\x $. The columns in the figure panel depict the tunneling, crossover and shuttle regimes from left to right, increasing value of $\chi$. In the tunneling regime, the stationary solution has the form of a Fock state centered in the origin of phase space, while in the shuttle regime it has the toroidal shape characteristic of a coherent state, which oscillates around the center of phase space. 

Quantum transport is usually described by a Markovian ME, which can be shorten in the Lindblad form as 
\begin{equation}\label{LindbladianTotal}
    \frac{d}{dt}\rho=\mathcal{L}\rho,\quad \mathcal{L}=\mathcal{L}_0+\mathcal{L}_1.
\end{equation}
The Lindbladian $\mathcal{L}$ is separated in two parts, with $\mathcal{L}_1=\sum_{k=1}^M \mathcal{J}_k$ describing $M$ different types of jump (tunneling) processes. In our case of interest \eqref{ME}, the jump refers to single electron tunneling, and is chosen as 
\begin{equation}\label{Jump}
\mathcal{L}_1\rho=\mathcal{J}\rho=\Gamma c\x e^{-x_0(a\x+a)/\lambda} \,\rho \, c e^{-x_0(a\x+a)/\lambda}
\end{equation}
for simplicity. That is, the jump describes the electron tunneling from the source to the island and describes an emission of a hole into the source. A similar picture can be obtained by considering the jump describing tunneling from the island to the drain, corresponding to an emission of an electron into the drain.

Using the interaction picture respect to $\mathcal{L}_0$, we can find a path integral formal solution to the ME in terms of conditioned density operators \cite{ETIM1978246,Carmichael,Zoller,Petruccione,Brandes2008}. They
describe the non-unitary time-evolution of an initial density
operator, which is interrupted by different types of quantum jumps. In the case of only two jumps, the conditioned density operator has a normalization factor called waiting time distribution (WTD). Restricted to the case of two jumps of the same type $\mathcal{J}$ \eqref{Jump}, occurring at times $t_1=0$ and $t_2=\tau$, the WTD is simply calculated as
\begin{equation}\label{WTD}
    w(\tau)=\frac{\tr(\mathcal{J}e^{\mathcal{L}_0\tau}\mathcal{J}\rho_0)}{\tr(\mathcal{J}\rho_0)},
\end{equation}
where we select the stationary solution $\mathcal{L}\rho_0=0$ as initial density operator in the path integral formalism. 
For more details on the explicit calculation of the WTD in the vectorized space, see Appendix \ref{App:Vectorization}, and particularly Eq.\eqref{WTDVectorized}.

According to the WTD definition \eqref{WTD}, it quantifies the conditional probability
distribution of two tunneling events separated by period of $\tau$. This definition is limited to the long-time limit, where the system reach a stationary state \cite{Cerrillo2022} and the time translational invariance is present \cite{Zhao2025}. Hence the mean waiting time can be defined as 
\begin{equation}\label{meantau}
    \la \tau\ra =\int_0^{\infty}\tau \,w(\tau)d\tau,
\end{equation}
and its standard deviation $\sigma_{\tau}=\sqrt{\la(\tau-\la \tau\ra)^2 \ra}$. Let us introduce the Fano factor $I=\sigma_{\tau}^2/\la \tau\ra^2$ as a magnitude to determine the signal-to-noise ratio \cite{Cerrillo2022}. The Fano factor is also presented in the literature as a quotient between the second (zero-frequency noise) and the first (mean value) cumulants of the current \cite{Novotny2004,Flindt2005}. This approach is based on full counting statistics, the stochastic theory of charge transfer through mesoscopic systems \cite{PhysRevB.67.085316,FlindtThesis}.

\begin{figure}[h]
	\centering
	\includegraphics[width=0.48\textwidth]{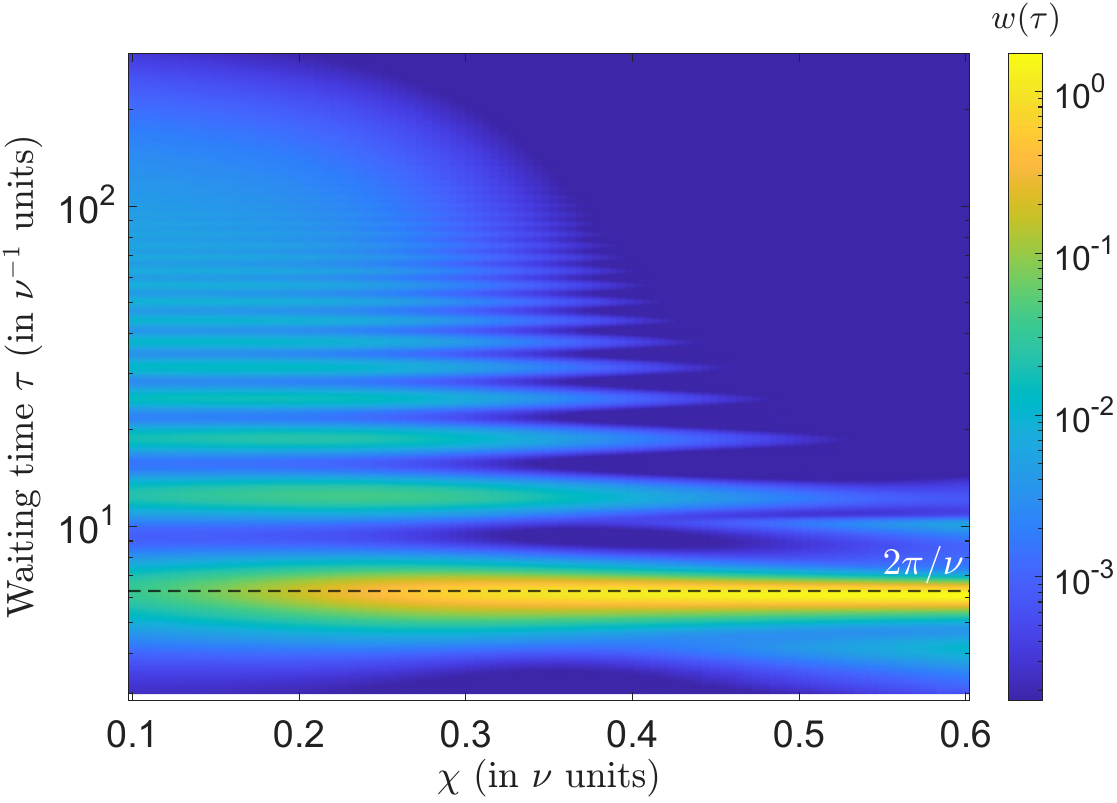}
	\caption{Heatmap of the waiting time distribution $w(\tau)$ as a function of the coupling $\chi$ (in $\nu$ units) and the waiting times (in $\nu^{-1}$ units). The WTD has been calculated in a 1000 elements time window of $\tau\in[0,300\nu^{-1}]$, in logarithmic scale to increase the point density for low waiting times. The horizontal dashed line identifies the period of the oscillator in the shuttle regime, $2\pi/\nu\simeq 6.28\nu^{-1}$. The parameters used are $\Gamma=0.01\nu$, $\kappa=0.05\nu$, $\lambda=2x_0$ and an oscillator basis of size $N=50$. }
	\label{Fig:WTD}
\end{figure}

\begin{figure}[h]
	\centering
	\includegraphics[width=0.48\textwidth]{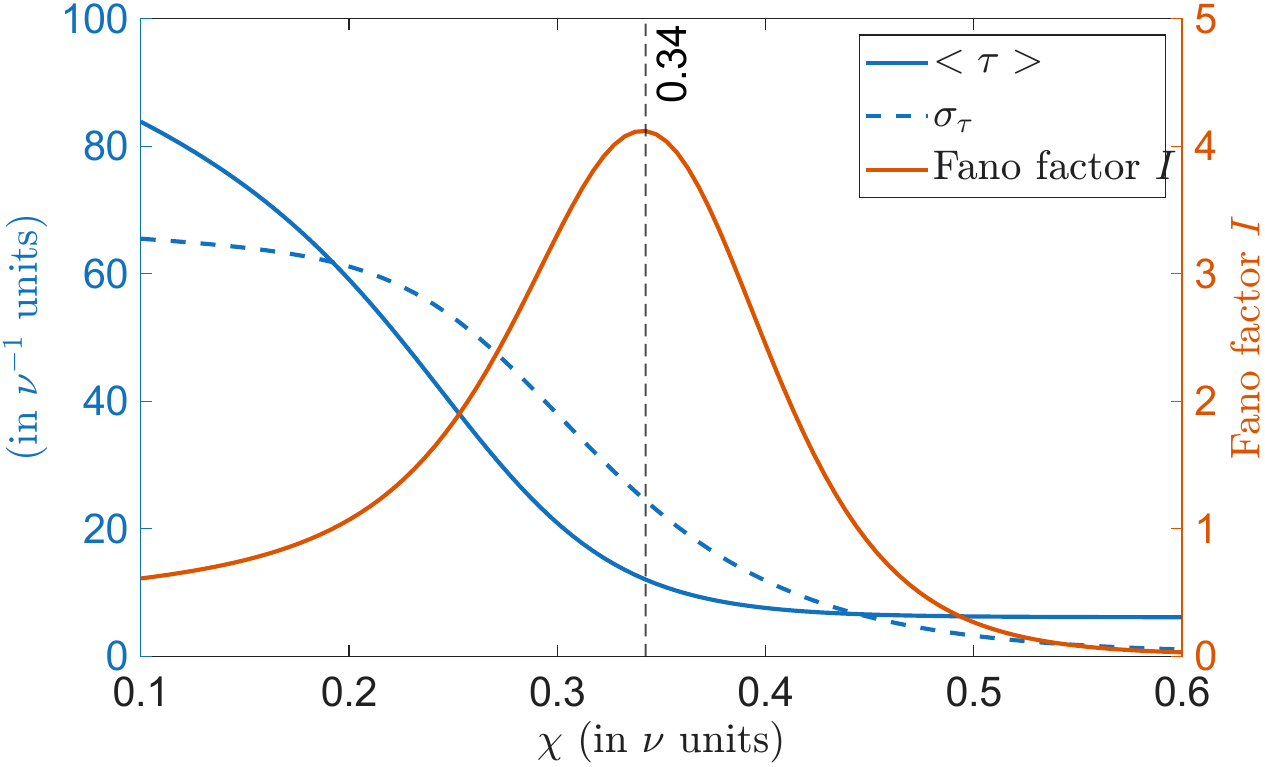}
	\caption{Mean waiting time (solid blue) and standard deviation (dashed) in $\nu^{-1}$ units at the left axis, and Fano factor (orange) at the right axis. Magnitudes are derived from the waiting time distribution of the electron shuttle \eqref{WTD}. The parameters used are $\Gamma=0.01\nu$, $\kappa=0.05\nu$, $\lambda=2x_0$ and an oscillator basis of size $N=50$. The WTD has been calculated in a 1000 elements time window of $\tau\in[0,300\nu^{-1}]$.}
	\label{Fig:Fano_factor_WTD}
\end{figure}

We plot the waiting time distribution \eqref{WTD} in Figure \ref{Fig:WTD}, as a function of the coupling $\chi$ (in $\nu$ units) and the waiting times (in $\nu^{-1}$ units). The parameters used are common along the figures, which are $\nu=1$s$^{-1}$, $\Gamma=0.01\nu$, $\kappa=0.05\nu$, $\lambda=2x_0$ and an oscillator basis of size $N=50$. The WTD has been calculated in a 1000 elements time window of $\tau\in[0,300\nu^{-1}]$, with logarithmic scale to increase the point density for low waiting times. This scaling becomes crucial for a non-variable time window, as the mean waiting time quickly decreases when reaching the shuttle regime, as we will show in Figure \ref{Fig:Fano_factor_WTD}. The WTD has been calculated in a 1000 elements time window of $\tau\in[0,300\nu^{-1}]$. On the left part of Figure \ref{Fig:WTD}, we see oscillatory pattern for the different vertical section of fixed $\chi$. That is, the WTD wave has two components, a long period envelope whose mean time decreases with $\chi$ and whose shape is typical in single-electron tunneling \cite{Cerrillo2022,donarini2024quantum,datta2005quantum}; and short period oscillations that only appear in the quantum version of the model \cite{Zhao2025}, not in the semiclassical approximation \cite{Cerrillo2022}. When increasing $\chi$, these oscillations shrink and the base frequency $\nu$ increasing, reaching its maximum on the right side of the figure for the shuttle regime. The horizontal dashed line identifies the period of the oscillator $2\pi/\nu\simeq 6.28\nu^{-1}$.

In addition, we display the mean waiting time \eqref{meantau} in Figure \ref{Fig:Fano_factor_WTD}, derived from the waiting time distribution of the electron shuttle \eqref{WTD}. The parameters used the same as in Figure \ref{Fig:Fano_factor_WTD}, and so does the waiting time window. The mean waiting time stagnates at $2\pi/\nu$ in the shuttle regime (high $\chi$), while it increases monotonically when moving to the tunneling regime (low $\chi$). In the same Figure \ref{Fig:Fano_factor_WTD}, we plot the standard deviation and the Fano factor for the same parameters. Both magnitudes display an inflection point and maximum, respectively, around $\chi=0.34\nu$, characterizing the crossover region previously depicted in Figure \ref{Fig:Husimi}. This is in agreement with the semiclassical version of the model \cite{Isacsson_2004}, where the Fano factor has a singularity (limit cycle) in the crossover between underdamped tunneling and shuttle regimes \cite{Cerrillo2022,Milburn2006}. 

\begin{figure}[h]
	\centering
	\includegraphics[width=0.48\textwidth]{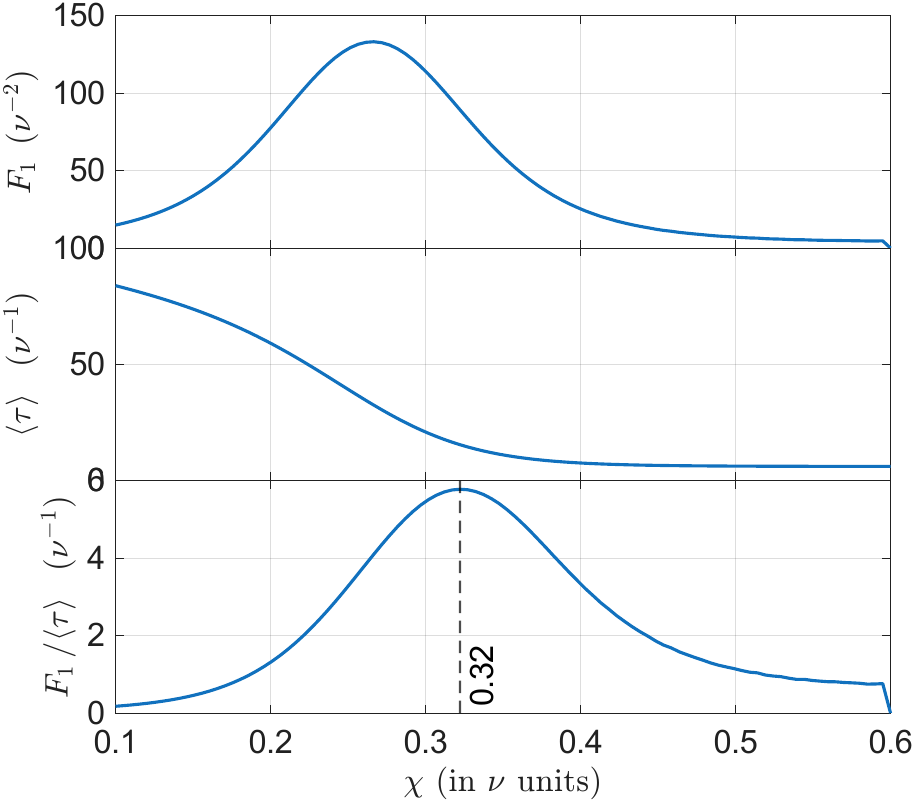}
	\caption{Classical Fisher Information of a single measurement \eqref{CFIeq} (top), mean waiting time \eqref{meantau} and CFI of multiple measurements divided by the measuring time $F(\chi)/T=F_1/\la\tau\ra$. Magnitudes are in time units ($\nu^{-1}$, except from the top in $\nu^{-2}$), and are derived from the waiting time distribution of the electron shuttle \eqref{WTD}. The parameters used are $\Gamma=0.01\nu$, $\kappa=0.05\nu$, $\lambda=2x_0$ and an oscillator basis of size $N=50$. The WTD has been calculated in a 1000 elements time window of $\tau\in[0,300\nu^{-1}]$.
    }
	\label{Fig:CFI}
\end{figure}

We have chosen the coupling  $\chi$, proportional to the electric field, as the parameter that we want to estimate. 
The electric field cannot be measured directly, so it is estimated from the results of measurements performed on $n$ identical copies of the output state, which is the stationary solution $\rho_0$ of the Lindbladian in our case. In particular, we measure the waiting time $\tau$ between two consecutive events, then compute the conditional probability or likelihood function $w(\tau|\chi)$ as the WTD \eqref{WTD}, and eventually estimate the electric field $\chi$. The achievable precision of the measurement is bounded by the Cram\'er-Rao bound
\begin{equation}
    \Delta\chi ^2\geq \frac{1}{F(\chi)}\,,
\end{equation}
where $F(\chi)=nF_1(\chi)$ is the classical Fisher information (CFI), defined as the Fisher information of one measurement $F_1(\chi)$ multiplied by the number $n$ of independent measurements on identical copies of the probe state. The function $F_1(\chi)$ is defined using the waiting time distribution as likelihood function \cite{10.1098/rsta.1922.0009}, that is 
\begin{equation}\label{CFIeq}
    F_1(\chi)=\int_0^\infty \left(\frac{\partial}{\partial \chi}\log w(\tau|\chi)\right)^2 w(\tau|\chi)\,d\tau\,
\end{equation}
as usually for unbiased estimators and independent measures.
In practice, the upper bound of the integral is truncated to a finite value $t_{\text{max}}$ when the WTD approaches asymptotically to 0. For continuously measured quantum systems, the number of measurements $n=T/\la \tau\ra$ is equivalent to the quotient of the total measured time $T$ and the measurements mean time $\la \tau\ra$, which is equivalent to the mean time between the tunneling events or the mean waiting time \eqref{meantau} \cite{PRXQuantum.5.020201,PhysRevLett.112.170401}. We are interested in the CFI for $n$ measurements in $T$ units, to make it independent of the measurement time, that is, we plot $F(\chi)/T=F_1(\chi)/\la\tau\ra$ at the bottom panel of Figure \ref{Fig:CFI}. Using the same parameters as in Figure \ref{Fig:Fano_factor_WTD}, we see how this quantity reaches its maximum around $\chi=0.32\nu$, detecting the crossover region close to the maximum of the Fano factor. At the top of the same figure we represent the CFI for a single measurement \eqref{CFIeq}, whose maximum is slightly moved to the left compared to the bottom panel.  

There is an upper bound for the CFI that requires maximizing over all possible measurements \cite{PhysRevLett.72.3439}, resulting in the quantum Fisher information (QFI) \cite{RevModPhys.90.035005,doi:10.1142/9789814338745_0015,10.1116/5.0007577,Denes_Petz_2002}. For a continuous-measurement metrology paradigm and a sufficiently large measuring time $T$, the QFI for the estimation of the parameter $\chi$ is given asymptotically by \cite{PRXQuantum.5.020201,PhysRevLett.112.170401},
\begin{equation}\label{QFIeq}
    F_Q[\rho_0]=-4T\,\tr\left[(\tfrac{1}{\hbar}\partial_\chi H)\mathcal{L}^+ (\{\rho_0,\tfrac{1}{\hbar}\partial_\chi H\})\right].
\end{equation}
Note that this expression is only valid for a system where the estimation parameter $\chi$ is encoded only in the Hamiltonian $H$, not in the jump superoperators of the dissipators (see eq.\eqref{Hamiltonian},\eqref{ME}). In addition, the derivative terms are reduced to $\tfrac{1}{\hbar}\partial_\chi H=c\x c(a\x +a)$ \eqref{Hamiltonian}. The symbol $\mathcal{L}^+$ represents the Drazin pseudoinverse \cite{Drazin01081958}, whose calculation requires diagonalizing the Lindbladian. However, we consider another method that is less computationally intensive and consists of solving a linear equation system, as we did with the WTD (see Appendix \ref{App:Vectorization}, and specially eqs.\eqref{WTDVectorized},\eqref{QFIvectorized},\eqref{QFIsystem}, for a detailed calculation). 

\begin{figure}[h]
	\centering
	\includegraphics[width=0.48\textwidth]{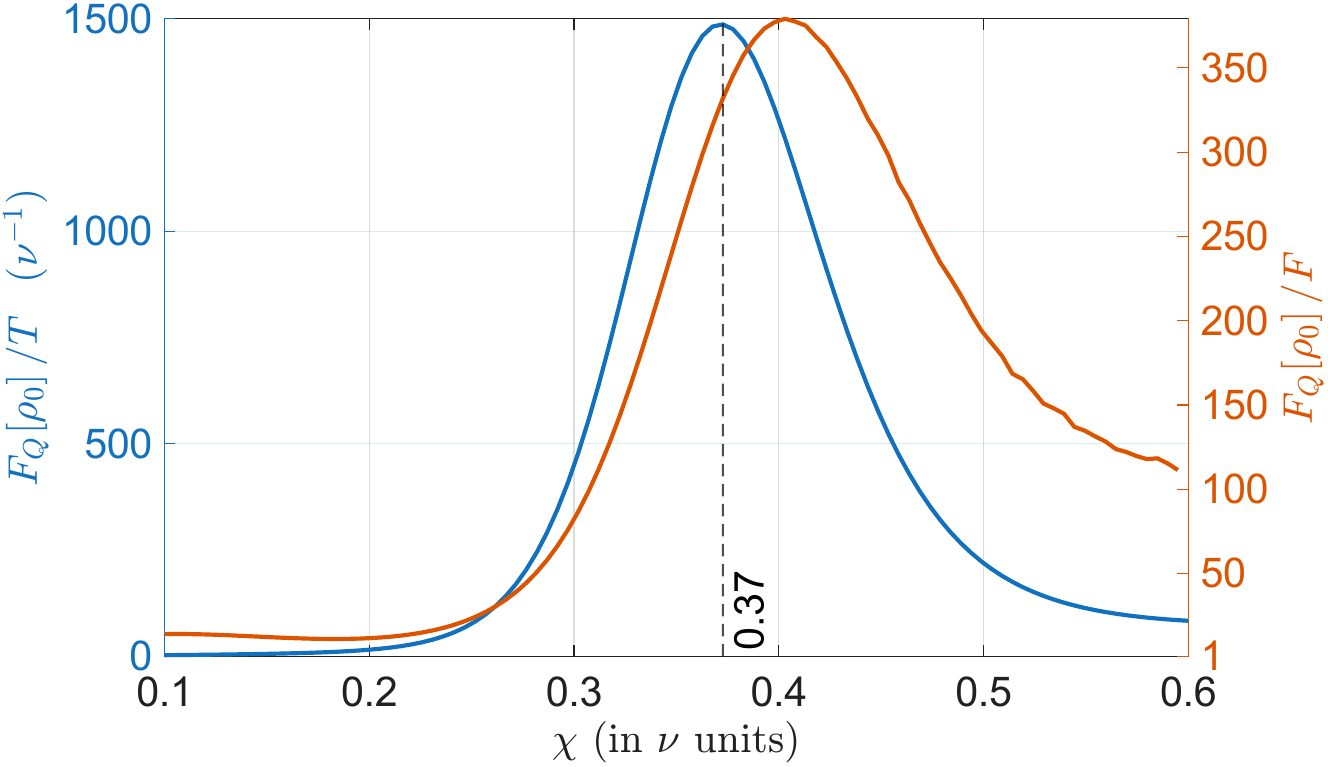}
	\caption{Quantum Fisher Information for a continuous measurement \eqref{QFIeq} divided by the measuring time $T$ (blue), and QFI divided by the CFI of Figure \ref{Fig:CFI} bottom panel (orange), both as function of the coupling $\chi$. The parameters used are  $\Gamma=0.01\nu$, $\kappa=0.05\nu$, $\lambda=2x_0$ and an oscillator basis of size $N=50$.}
	\label{Fig:QFI}
\end{figure}

In the blue line of Figure \ref{Fig:QFI}, we represent the QFI of eq.\eqref{QFIeq} divided by $T$, so that it is independent of the measuring time. As the Drazin pseudoinverse has $\nu^{-1}$ units and $\tfrac{1}{\hbar}\partial_\chi H$ is dimensionless, the QFI $F_Q[\rho_0]/T$ in the plot has $\nu^{-1}$ units, as the CFI in Figure \ref{Fig:CFI}. Although the QFI presents a peak for a slightly higher value $\chi=0.37$ than the CFI plot, the crossover regime is detected in the same region of the coupling parameter. To compare QFI and CFI, we plot in the orange line the quotient between both $F_Q[\rho_0]/F=\tfrac{F_Q[\rho_0]/T}{F_1/\la \tau\ra}$, which is equivalent to dividing the blue line in Figure \ref{Fig:QFI} by the bottom panel of Figure \ref{Fig:CFI}. As the QFI is an upper bound for the CFI \cite{RevModPhys.90.035005,HELSTROM1967101}, the orange line in Figure \ref{Fig:QFI} stays above 1 for all $\chi$. The quotient achieves maximum values around the crossover, when the semiclassical version of the model cannot reproduce the effect of its quantum counterpart.

\section{Comparison to trapped ions}\label{comparison} 

Estimating weak electric fields is a central task in quantum sensing, and a variety of quantum platforms have been proposed and experimentally realized for this purpose. In this work, we focus on an electron shuttle, in which the electric field couples directly to the mechanical motion of a charged nanostructure. To assess the performance of this approach, it is useful to compare it with alternative sensing architectures. 

Among the most successful quantum sensing platforms are trapped ions. Although trapped-ion experiments are often formulated as force- or displacement sensing protocols rather than electric-field measurements, the displacement of a charged ion is ultimately induced by an external electric field. As a result, the sensitivity achieved in displacement sensing experiments can be directly translated into electric-field sensitivity. This observation enables a quantitative comparison of the performance of electron shuttles and trapped ions within a common metrological framework. 

In particular, trapped-ion displacement sensing has been investigated both theoretically and experimentally in several works, including the protocol of Wolf \textit{et al.}~\cite{Wolf2019} as well as related studies on quantum-enhanced displacement estimation and force sensing~\cite{Cerrillo2021,Delakouras2024}. Since both trapped ions and electron shuttles probe an external electric field through the induced motion of a charged oscillator, the corresponding Fisher information can be expressed in terms of the same physical parameter, namely the electric field amplitude. We therefore compare the attainable electric-field sensitivities of both platforms by translating the Fisher information associated with the measured displacement into the Fisher information for electric-field estimation. Beyond benchmarking the present shuttle implementation, this comparison allows us to identify the physical parameters that should be adjusted to approach the sensitivity achieved with trapped ions.

In the shuttle, the parameter to be estimated is $\chi$, which is proportional to the electric field $E$ according to $\hbar \chi = e x_0 E$. Consequently, the Fisher information associated with the electric field is \begin{equation} 
F(E)=\left(\frac{e x_0}{\hbar}\right)^2F(\chi). 
\end{equation}
In trapped-ion force sensing experiments the displacement is typically parameterized by the total coherent displacement amplitude $\alpha$. Because the accumulated displacement scales linearly with the interrogation time $t_i$, the corresponding Fisher information for trapped ions $F_i$ can be written as \begin{equation} 
F_i(E)=\left(\frac{e x_{0,i} t_i}{\hbar}\right)^2F_i(\alpha),
\end{equation} 
with $x_{0,i}=\sqrt{\hbar/(2m_i\nu_i)}$ the zero point motion of an ion of mass $m_i$ in a trap of frequency $\nu_i$. Therefore, the ratio of the two sensitivities becomes 
\begin{equation}\label{FisherComparison} 
\frac{F(E)}{F_i(E)} = \frac{m_i\nu_i }{m\nu }  \frac{F_1(\chi)}{\la \tau\ra  t_i F_i(\alpha)}.
\end{equation} 
Equation~(\ref{FisherComparison}) shows that, for current shuttle realizations, the primary limitation is the mechanical oscillator's large effective mass relative to that of a single trapped ion. However, the shuttle produces a continuous sequence of tunneling events, allowing information to be accumulated from successive waiting-time measurements. The two main routes to improve electric-field sensitivity are therefore to reduce $m\nu$ and to increase the Fisher information rate extracted from the transport record.


Therefore, achieving the trapped-ion benchmark considered here requires either higher Fisher information for the shuttle, a lighter oscillator, or both. Fisher information can be increased by operating around the crossover, where it peaks, and by using measurements that recover more of quantum Fisher information than waiting-time measurements alone. The mechanical limitation can be mitigated by using atomic-scale quantum dots or other nanomechanical devices with masses below those of current shuttle realizations. Such systems could retain the enhanced response around the crossover while increasing the zero-point motion. The comparison therefore identifies three routes to improve shuttle sensing: reducing $m\nu$, operating around the crossover, and extracting more information from the transport record.

\section{Conclusions}\label{conclusion}

We have investigated electron waiting times in a noisy quantum shuttle and compared them with the total parameter sensitivity in the stationary island-oscillator state. The Husimi distributions demonstrate the transition from a localized tunneling regime to a shuttling regime. The waiting time distributions reveal significant changes in transport statistics as the electromechanical coupling increases. This reorganization is also reflected in the first two moments of the waiting time distribution. The mean waiting time decreases as mechanically assisted transport becomes established, whereas the Fano factor shows a pronounced enhancement in the crossover region before fluctuations are suppressed in the more regular shuttling regime. The crossover is therefore the noisiest operating region in relative terms, even though it is also the region where the dynamics respond most strongly to changes in the electromechanical coupling. Treating the coupling strength, and hence the applied electric field, as an unknown parameter, we used classical and quantum Fisher information to quantify the response. Under the same resource normalization, the waiting-time Fisher information remains below the quantum Fisher information across all regimes studied. Hence, monitoring the time intervals between tunneling events does not fully recover information about the joint electronic and mechanical degrees of freedom. The comparison links noise to metrological sensitivity, revealing increased fluctuations that signal transport dynamics sensitive to electromechanical coupling. The Fano factor and Fisher information peak in the same region, but noise and information aren't identical; they may peak at different parameters. Large variance alone doesn't guarantee optimal estimation; what matters is how the probability distribution or quantum state changes with the parameter.

These findings establish that waiting-time measurements provide a transparent, relevant sensing protocol, while quantum Fisher information offers a comprehensive benchmark for identifying the best operating point or comparing measurement strategies. For evaluating the device's sensing potential, determining optimal conditions, or comparing detection schemes, quantum Fisher information is the best figure of merit. Future work could measure them near the quantum limit, compare them at the level of monitored trajectories, and assess robustness to temperature, tunneling asymmetry, detector inefficiency, and non-Markovian environments.

\section*{Acknowledgments} 
AM, ALG, EH and JC acknowledge support from grant CNS2023-144994 funded by MICIU/AEI/10.13039/201100011033 and "ERDF/EU" and from European Union project C-QuENS (Grant No. 101135359).
AM additionally acknowledges support of the Spanish Ministry of Science through the project PID2022-138144NB-I00. 

\appendix 

\section{Approximations in the master equation}\label{App:Approximations}

In the following, we present a series of approximations necessary to obtain the simplified ME in \eqref{ME}. Firstly, it is enlightening to show the full expression of the total Hamiltonian \eqref{HamTotal} according to the references \cite{Milburn2006,Zhao2025},
\begin{align}
    \begin{array}{lllr}
		H_\text{T}=&\hbar\omega_\text{I}c\x c+ U_c (c\x c)^2+ \hbar\nu a\x a - \hbar\chi c\x c (a\x + a)& (H_\text{S})\\[10pt]
    &+\sum_k(\hbar \omega_{sk}c_k\x c_k+\hbar\omega_{dk}d_k\x d_k)& (H_\text{L})\\[10pt]
    &+\sum_k(T_{sk}e^{-\hat{x}/\lambda}c_kc\x+\text{H.c.})&(H_\text{tun})\\[10pt]
    &+\sum_k(T_{dk}e^{\hat{x}/\lambda}d_kc\x+\text{H.c.})&\\[10pt]
    &+\sum_p\Big[\hbar\omega_pb_p\x b_p+g(a\x b_p+a b_p\x )\Big]& (H_\text{B})
	\end{array}
\end{align}
Comparing with eq.\eqref{Hamiltonian}, in the system Hamiltonian $H_\text{S}$, the strong Coulomb blockade regime ensures that no more than one electron occupies the island at any time, so that the number operator fulfills $(c\x c)^2=c\x c$ and we can define a single effective energy level $\varepsilon_{I}=\hbar\omega_\text{I}+U_c$.
The leads Hamiltonian $H_{\text{L}}$ describes the electrostatic
energy of an ensemble of non-interacting electrons $c_k$ and $d_k$, described by the Fermi distribution. The tunneling Hamiltonian $H_{\text{tun}}$ couples the island, the leads, and the oscillator via the exponential of the position operator $\hat{x}=\sqrt{\frac{\hbar}{2 m \nu}}(a+a\x)$. The bath Hamiltonian $H_\text{B}$ couples the oscillator to a dissipative Ohmic heat bath (bosonic $b_p$) in the rotating wave approximation.

The interaction of the system $H_\text{S}$ \eqref{Hamiltonian} with the bosonic heat bath $H_\text{B}$ has decoherence effects, so that the initial quantum state, in absence of driving force, evolves to a thermal stationary distribution (diagonal density matrix) at long time. This relaxation towards canonical equilibrium, the translational invariance that shows the damping term, and the positivity of the density matrix are incompatible with a Markovian damping kernel \cite{Novotny2003}. Therefore, we weaken the positivity as in \cite{10.1063/1.474887}. Nevertheless, we checked numerically that the positivity is only broken for large values of the damping $\kappa$ out of the shuttle regime, in accordance with \cite{Novotny2003,DonariniThesis}. As an additional approximation, we set the damping rate $\kappa$ constant, fixing the density of states of the phonon bath and the oscillator-bath coupling \cite{DonariniThesis}.

In contrast to the weak coupling (Born approximation) with the thermal bath, the tunneling coupling with the leads is comparable to the oscillator frequency. Hence, in the shuttle regime, the time scale of the electrical dynamics is comparable with the period of the mechanical oscillations in the system. This is why we have considered an arbitrary coupling to the leads approach \cite{PhysRevB.53.15932}. Firstly, the island-leads coupling in the ME \eqref{ME} is computed with the strong Coulomb blockade. This means that once the dot is charged with an electron of specific spin, only that species can tunnel out. This approximation is equivalent to renormalize the injection rate and consider spin-less non-interacting particles in the island \cite{DonariniThesis,PhysRevB.60.14318}.

To obtain a Markov master equation for the island-oscillator system, we trace out the degrees of
freedom in the leads and use the Born-Markov approximation \cite{Petruccione,PhysRevB.70.075303}. That is, we
assume the vibrational frequency of the oscillator is slow
compared to bath relaxation time scales. The derived ME is \cite{Milburn2006,DonariniThesis,Zhao2025}
\begin{align}
    \frac{d}{dt}\rho=&\,-i\nu\left[a\x a,\rho \right]+i\chi\left[c\x c(a\x+a),\rho \right] \nonumber\\
    &+ \Gamma_Lf(\varepsilon_\text{I}-\mu_L)\, \Dc\left[c\x e^{- x_0(a\x + a)/\lambda }\right]\rho\nonumber\\
    &+ \Gamma_L(1-f(\varepsilon_\text{I}-\mu_L))\, \Dc\left[c\,e^{- x_0(a\x + a)/\lambda }\right]\rho\nonumber\\
    &+ \Gamma_Rf(\varepsilon_\text{I}-\mu_L)\, \Dc\left[c\x e^{x_0(a\x + a)/\lambda }\right]\rho\nonumber\\
    &+ \Gamma_R(1-f(\varepsilon_\text{I}-\mu_L))\, \Dc\left[c\,e^{ x_0(a\x + a)/\lambda }\right]\rho\nonumber\\
    &+ \kappa (\bar{n}_p+1) \,\Dc\left[a\right]\rho+\kappa \bar{n}_p \,\Dc\left[a\x\right]\rho\,.
\end{align}
We consider leads temperatures much smaller than their Fermi energy, so we can approximate their Fermi distributions $f(\varepsilon_\text{I}-\mu_{L,R})$ by step functions. The chemical potential of the left (right) lead is assumed much higher (lower) than the dot energy level $\mu_L\gg\varepsilon_\text{I}\gg \mu_R$. We also consider the high bias approximation, that is the tunneling amplitude and density of states of the leads are constant, making constant the injection and ejection rates $\Gamma=\Gamma_L=\Gamma_R$ (we additionally set both equal) \cite{PhysRevB.70.075303}. We have also disregarded cotunneling and higher order scattering events. The local thermodynamic equilibrium in the leads makes that any correlation, between the electrons in the leads and in the
island, rapidly decays to zero, as a result of the tunneling interaction and on time scales relevant for the system
dynamics \cite{PhysRevB.70.075303}. The Bose-Einstein distribution $\bar{n}_p$ of the bath, in thermal equilibrium, is also approximated to 0 \cite{Novotny2004}. All together leads to the reduced ME \eqref{ME}.

The last approximation deriving the ME driving terms $\Dc\left[c^{(\dagger)}e^{ \pm(a\x + a)/\lambda }\right]\rho$  (island-oscillator-leads coupling), is to assume that the position of the system energy levels should be contained in the transport window open between the chemical potentials of the leads. To be precise, only a finite number of mechanical excitations are involved in the dynamics of the system, which means a truncation of the oscillator basis. 
Even in the presence of the bath, this assumption is numerically accomplished in the cases of study,
being the violation of this condition incongruent with the validity of the ME. 

\section{Vectorization of the island-oscillator system}\label{App:Vectorization}


The density matrices solution of the master equation \eqref{ME} belong to the tensor product Hilbert space of the island-oscillator $\mathcal{H}_I\otimes \mathcal{H}_O$. The oscillator space is truncated to dimension $N$ for numerical calculations, hence the density matrix has a size $2N\times 2N$. In order to solve the ME, we perform a vectorization to express the matrix multiplication as a linear transformation. That is, $A\in\mathcal{M}_{2N,2N}(\mathbb{C})$ matrices are transformed into $\text{vec}(A)\in\mathcal{M}_{4N^2,1}(\mathbb{C})$ column vectors, and the following expressions are used
\begin{align}
    \text{vec}(ABC)=&\,(C^T\otimes A)\text{vec}(B),\\
    \text{vec}(AB)=&\,(I_{2N}\otimes A)\text{vec}(B)=(B^T\otimes I_{2N})\text{vec}(A),\\
    \tr(A\x B)=&\,\text{vec}(A) \x \text{vec}(B),\label{traceVec}
\end{align}
for all $A\in\mathcal{M}_{2N,2N}(\mathbb{C})$, where $I_{2N}=I_{2}\otimes I_{N}$ is the identity matrix of the island-oscillator space. The commutator and dissipator terms in the ME \eqref{ME} are transformed into
\begin{align}
    \text{vec}([A,\rho])=&\,(I_{2N}\otimes A-A^T\otimes I_{2N})\text{vec}(\rho), \label{Vectorization1}\\[4pt]
    \text{vec}(\mathcal{D}[A]\rho)=&\,(A^*\otimes A)\text{vec}(\rho) \label{Vectorization2}\\
    &\,+\frac{1}{2}(I_{2N}\otimes A+A^T\otimes I_{2N})\text{vec}(\rho) \nonumber.
\end{align}

The problem of calculating the stationary solution of the ME $\mathcal{L}\rho_{0}=0$ is reduced to solving a system of linear equations $\text{vec}(\mathcal{L}\rho_{0})=L\,\text{vec}(\rho_0)=0$,  where $L$ is a $4N^2\times 4N^2$ matrix obtained by applying Eqs.\eqref{Vectorization1},\eqref{Vectorization2} to the ME \eqref{ME}. The linear system dimensions increase by one because of the extra normalization condition $\tr(\rho)=\text{vec}(I_{2N})^T \text{vec}(\rho)=1$ as in references \cite{Brandes2008,Flindt2005}. The solution of the system is computed using the MatLab\textsuperscript{\textcopyright} \textit{mldivide} routine, that combines brute force and LU solver. It is worth mentioning that other works use the Arnoldi scheme (Krylov spaces, singular value decomposition and preconditioning) to reduce the dimensionality and accelerate the computations \cite{Flindt2004,DonariniThesis,Novotny2003,Novotny2004}.

The jump operator chosen in Eq.\eqref{Jump} has a vectorized form
\begin{align}
    \text{vec}({\mathcal{J}}\rho)=&\,\Gamma (c\x e^{-x_0(a\x+a)/\lambda})\otimes (c\x e^{-x_0(a\x+a)/\lambda})\text{vec}(\rho)\nonumber\\
    =&\,J\, \text{vec}(\rho),
\end{align}
where $J$ is the $4N^2\times 4N^2$ superoperator form. The waiting time distribution in Eq.\eqref{WTD} is also vectorized as
\begin{equation}\label{WTDVectorized}
    w(\tau)=\frac{\text{vec}(I_{2N})^T J Ve^{D\tau}V^{-1}J\text{vec}(\rho_0)}{\text{vec}(I_{2N})^T \text{vec}(\rho_0)}.
\end{equation}
The symbols represent $V$ and $D$ represent the eigenvector and eigenvalues matrices derived from the Lindbladian $L_0$ after the vectorization, that is, $\text{vec}(\mathcal{L}_0\rho)=L_0\text{vec}(\rho)$ (see Eq.\eqref{LindbladianTotal}). Note that, instead of computing the inverse matrix $V^{-1}$, it is easier to solve the equation system $VX=J\text{vec}(\rho_0)$. In addition, the WTD is computed for a finite number of points in the $\tau$-domain as $w\xrightarrow{ \tau \to \infty }0$, that is, we truncate up to certain value $\tau=t_{\text{max}}$ coinciding with the upper bound in the CFI integral \eqref{CFIeq}.

The QFI expression \eqref{QFIeq} can also be vectorized to simplify its calculation. Firstly, we use the eq.\eqref{traceVec}, so that the QFI transforms into
\be\label{QFIvectorized}
-4T\,\tr\left[(\tfrac{1}{\hbar}\partial_\chi H)\mathcal{L}^+ B\right]
=\text{vec}(\tfrac{1}{\hbar}\partial_\chi H^{\dagger})^\dagger\,\text{vec}(\mathcal{L}^+ B)\,,
\ee
where we have defined $B=\{\rho_0,\tfrac{1}{\hbar}\partial_\chi H\}$ to simplify the notation. Instead of diagonalizing the Lindbladian and computing $\mathcal{L}^+$, we define $X=\mathcal{L}^+ B$ and multiply this expression by $\mathcal{L}$ to obtain $\mathcal{L}X=\mathcal{L}\mathcal{L}^+ B$. The product $\mathcal{L}\mathcal{L}^+$ reduces to $\mathbb{1}-\text{vec}(\rho_0)\text{vec}(I_{2N})^T$ when vectorizing, which is the identity on the subspace complementary to that defined by $\text{vec}(\rho_0)\text{vec}(I_{2N})^T$ \cite{PRXQuantum.5.020201}. Then, we can write the equation system
\be\label{QFIsystem}
L\,\text{vec}(X)=(\mathbb{1}-\text{vec}(\rho_0)\text{vec}(I_{2N})^T)\text{vec}(B)\,
\ee
where $L$ comes from the vectorization of $\text{vec}(\mathcal{L}\rho)=L\,\text{vec}(\rho)$. Eventually, the solution of the system $\text{vec}(X)$ is equivalent to the term $\text{vec}(\mathcal{L}^+ B)$ by definition, and the QFI can be computed in the vectorized version of eq.\eqref{QFIvectorized}. 

\bibliography{bibliografia.bib}

\end{document}